\documentclass[10pt]{article}

\usepackage{amssymb}
\usepackage{latexsym}
\usepackage{amsmath}  
\usepackage{amscd}

\numberwithin{equation}{section}

\newcommand{\bR}{{\mathbb R}}

\newcommand{\bC}{{\mathbb C}}

\newcommand{\kB}{{\mathcal B}}

\newcommand{\gotH}{{\mathfrak H}}
\newcommand{\goth}{{\mathfrak h}}

\newcommand{\gotk}{{\mathfrak k}}

\newcommand{\ga}{{\alpha}}

\newcommand{\gD}{{\Delta}}

\newcommand{\gl}{{\lambda}}

\newcommand{\go}{{\omega}}

\newcommand{\gt}{{\tau}}

\newcommand{\slim}{\,\mbox{\rm s-}\hspace{-2pt} \lim}

\newcommand{\imag}{{\Im{\mathrm m\,}}}
\newcommand{\dom}{{\mathrm{dom}}}

\newtheorem{theo}{Theorem}[section]
\newtheorem{prop}[theo]{Proposition}
\newtheorem{lem}[theo]{Lemma}
\newtheorem{cor}[theo]{Corollary}

\newtheorem{defi}[theo]{Definition}
\newtheorem{rem}[theo]{Remark}

\newtheorem{conj}[theo]{Conjecture}

\newcommand{\ba}{\begin{array}}
\newcommand{\ea}{\end{array}}
\newcommand{\bea}{\begin{eqnarray}}
\newcommand{\eea}{\end{eqnarray}}
\newcommand{\bead}{\begin{eqnarray*}}
\newcommand{\eead}{\end{eqnarray*}}
\newcommand{\be}{\begin{equation}}
\newcommand{\ee}{\end{equation}}
\newcommand{\bed}{\begin{displaymath}}
\newcommand{\eed}{\end{displaymath}}
\newcommand{\bl}{\begin{lem}}
\newcommand{\el}{\end{lem}}
\newcommand{\bp}{\begin{prop}}
\newcommand{\ep}{\end{prop}}
\newcommand{\bt}{\begin{theo}}
\newcommand{\et}{\end{theo}}
\newcommand{\Label}{\label}
\newcommand{\bc}{\begin{cor}}
\newcommand{\ec}{\end{cor}}
\newcommand{\la}{\Label}

\newcommand{\br}{\begin{rem}}
\newcommand{\er}{\end{rem}}
\newcommand{\bd}{\begin{defi}}
\newcommand{\ed}{\end{defi}}

\newenvironment{proof}%
{\begin{sloppypar}\noindent{\bf Proof.}}%
{\hspace*{\fill}$\square$\end{sloppypar}}

\begin{document}

\title{Zeno product formula revisited}
\author{Pavel Exner$^{a}$, Takashi Ichinose$^{b}$, Hagen Neidhardt$^{c}$, \\and\\
Valentin A. Zagrebnov$^{d}$}
\date{\today}
\maketitle \maketitle

\begin{quote}
{\small \em a) Department of Theoretical Physics, NPI, Academy of
Sciences, \mbox{CZ-25068} \v Re\v z, and Doppler Institute, Czech
Technical University, B\v{r}ehov{\'a}~7,
CZ-11519 Prague, Czech Republic \\
b) Department of Mathematics, Faculty of Science, Kanazawa
University, Kanazawa 920-1192,
Japan\\
c) Weierstra{\ss}-Institut f\"ur Angewandte Analysis und
Stochastik,
Mohrenstr. 39, D-10117 Berlin, Germany \\
d) D\'{e}partement de Physique, Universit\'{e} de la
M\'{e}diterran\'{e}e (Aix-Marseille II) and  Centre de Physique
Th\'{e}orique, CNRS, Luminy Case 907, F-13288
Marseille Cedex 9, France \\
\rm exner@ujf.cas.cz, ichinose@kenroku.kanazawa-u.ac.jp,\\ 
neidhard@wias-berlin.de, zagrebnov@cpt.univ-mrs.fr} \vspace{8mm}

\noindent {\small Abstract:}
We introduce a new product formula which combines an
orthogonal projection with a complex function of a non-negative
operator. Under certain assumptions on the complex function 
the strong convergence of the product formula is shown. Under more
restrictive assumptions even operator-norm convergence is verified.
The mentioned formula can be used to describe
Zeno dynamics in the situation when the usual non-decay
measurement is replaced by a particular generalized observables in
the sense of Davies.
\end{quote}

\renewcommand{\thefootnote}{\fnsymbol{footnote}}

\setcounter{footnote}{0}
\renewcommand{\thefootnote}{\arabic{footnote}}

\maketitle

\renewcommand{\thefootnote}{\fnsymbol{footnote}} \setcounter{footnote}{0} %
\renewcommand{\thefootnote}{\arabic{footnote}}

\section{Introduction}

Product formul{\ae} are a traditional tool in various branches of
mathematics; their use dates back to the time of Sophus Lie. Such formul{\ae}
are often of the form 
\be\la{0.1}
\slim_{n\to\infty}\left(e^{-itA/n}e^{-itB/n}\right)^n = e^{-itC},
\quad C := A + B, \quad t \in \bR,
\ee
where $A$ and $B$ are bounded operators
on some separable Hilbert space $\gotH$ and $\slim$ stands for the
strong operator topology. A natural generalization to
unbounded self-adjoint operators $A$ and $B$ is due to Trotter \cite{Tr1,Tr2}
who showed that the limit exists and is equal to $e^{-itC}$, $t \in
\bR$, if the operator $C$,
\bed
Cf := Af + Bf, \quad f \in \dom(C) := \dom(A) \cap \dom(B),
\eed
is essentially self-adjoint. In  \cite{Ka1,Ka2}
Kato focused his interest to products of the type
\be\la{0.2}
\slim_{n\to\infty}\left(f(tA/n)g(tB/n)\right)^n\;\mbox{and}\;
\slim_{n\to\infty}\left(g(tB/n)^{1/2}f(tA/n)g(tB/n)^{1/2}\right)^n
\ee
where $A,B$ are non-negative self-adjoint operators and $f,g$
are now so-called Kato functions. Recall that a Borel measurable
function $f(\cdot): [0,\infty) \longrightarrow \bR$ is usually called a
Kato function if the conditions
\bed
0 \le f(x) \le 1, \quad f(0) = 1, \quad f'(+0) = -1,
\eed
are satisfied. Under these conditions he was able to show that the limits \eqref{0.2} exist and are
equal to $e^{-tC}$, $t \in \bR$,  where $C$ is the form sum of $A$ and
$B$. Notice $f(x) = e^{-x}$ is a Kato-functions which yields
the well-known Trotter-Kato product formul{\ae}
\bed
\slim_{n\to\infty}\left(e^{-tA/n}e^{-tB/n}\right)^n  = 
\slim_{n\to\infty}\left(e^{-tB/2n}e^{-tA/n}e^{-tB/2n}\right)^n = e^{-tC},
\quad t \ge 0.
\eed
Both products \eqref{0.1} and \eqref{0.2} are very useful and admit
applications to functional
integration, quantum statistical physics and other
parts of physics -- see, e.g., \cite[Chap.~V]{Ex}, \cite{Za} and
references therein. 
The last decade brought a progress in understanding of
the convergence properties of such formul{\ae} in operator-norm and
trace-class topology, for a review of
these results we refer to the monograph \cite{Za}.

In the last few years we have witnessed a surge of interest to another
type of product formul{\ae}, namely those where the left-hand side is
replaced by an expression of the type
\be\la{0.3}
\slim_{n\to\infty} \left(Pe^{-itH/n}P\right)^n
\ee
where $H$ is a self-adjoint operator on some separable Hilbert space
$\gotH$ and $P$ is a orthogonal projection on some closed subspace $\goth
\subseteq \gotH$. 
Products of such type are motivated by the ``quantum
Zeno effect'' (QZE). We call them therefore  Zeno product
formul{\ae}. The fact that the limit, if it exists, may be unitary on
$\goth$  is a venerable problem known
already to Alan Turing and formulated in the usual decay context
of quantum mechanics for the first time by Beskow and Nilsson \cite{BN}: frequent
measurements can slow down a decay of an unstable system, or even
fully stop it in the limit of infinite measurement frequency. The
effect was analyzed mathematically by Friedman \cite{Fr} but
became popular only after the authors of \cite{MS} invented the
above stated name. Recent interest is motivated mainly by the fact
that now the effect is within experimental reach; an up-to-date
bibliography can be found, e.g., in \cite{FMP} or \cite{Sch}.
The physical interpretation of this formula can be given in the
context of particle decay, cf. \cite[Chap.~II]{Ex}. The
unstable system is characterized by a projection $P$ to a subspace
$\goth$  of the state Hilbert space $\gotH$ of a larger, closed system,
the dynamics of which is governed by a self-adjoint Hamiltonian
$H$. Repeating the non-decay measurement experiment with the period $t/n$, we
can describe the time evolution over the interval $[0,t]$ of a
state originally in the subspace $P\gotH$ by the interlaced
product $(P e^{-itH/n}P)^n$; the question is how this operator
will behave as $n\to\infty$. 

In \cite[Theorem 1]{MS} it was shown that if 
the limit \eqref{0.3} exists and there is a conjugation $J$ commuting with
$P$ and $H$, then the Zeno product formula defines a unitary group on the
subspace $\goth$. Another simple example shows that this result is not
valid generally: the limit \eqref{0.3}  may exist without defining a unitary group.
Let $\gotH = L^2(\bR)$ and $H$ be the momentum operator, i.e $H =
-i\partial_x$ and let $P = P_{[a,b]}$ be the orthogonal projection on some subspace
$\goth = L^2([a,b])$, $[a,b] \subseteq \bR$. A straightforward
calculation shows that
\bed
P_{[a,b]}e^{-isH}P_{[a,b]} = P_{[a,b]}P_{[a+s,b+s]}e^{-isH}P_{[a,b]},
\quad s\in \bR.
\eed
Therefore, we get 
\be\la{1.31}
T(t) := \slim_{n\to\infty} \left(Pe^{-itH/n}P\right)^n = 
Pe^{-itH}\upharpoonright \goth, \quad t \in \bR,
\ee
which is neither unitary nor it satisfies the group property but
defines a contraction semigroup for $t \ge 0$. 
This example is covered by the following more general one. 
Let $H$ be the minimal self-adjoint dilation of a 
maximal dissipative operator $K$ defined on the subspace $\goth$.
Since by definition of self-adjoint dilations, cf. \cite{FN},
\bed
P e^{-isH}P := e^{-isK} = P e^{-isH}\upharpoonright \goth , \quad s \ge 0,
\eed
we find
\bed
\slim_{n\to\infty} \left(Pe^{-itH/n}P\right)^n = e^{-itK}, \quad t
\ge 0.
\eed
>From now on the strong convergence in the product formula is considered
only on $\goth$. 
Further examples can be found in \cite{MaS}.
However, in all of them the non-unitarity of the limit is related to
the fact that $H$ is not semibounded.  
So we restrict ourself  in the following to the case that $H$ is
semi-bounded from below; it is clear that without loss of generality
we may assume that $H$ is non-negative.

It has to be stressed that the last mentioned assumption does not ensure the
existence of the limit \eqref{0.3}. Indeed, if $\dom(\sqrt{H}) \cap
\goth$ is not dense in $\goth$, then it
can happen that the left-hand side in the Zeno product formula does not converge, cf.
\cite[Rem.~2.4.9]{Ex} or \cite{MaS}.

With these facts in mind we assume in the following that $H$ is a non-negative
self-adjoint operator such that $\dom(\sqrt{H}) \cap \goth$ is dense
in $\goth$. Under these assumptions we claim that a ``natural'' candidate for the limit of the Zeno
product formula \eqref{0.3} is the unitary group $e^{-itK}$, $t \in \bR$, on
$\goth$, generated by the non-negative self-adjoint operator $K$ associated with the
closed sesquilinear form $\gotk$,
\be\la{1.5}
\gotk(f,g) := (\sqrt{H}f,\sqrt{H}g), \quad f,g \in \dom(\gotk) = \dom(\sqrt{H})
\cap \goth \subseteq \goth.
\ee
The claim rests upon the paper
\cite[Theorem 2.1]{EI} where it is shown that 
\be\la{1.5a}
\lim_{n\to\infty}\int^T_0 \|\left(Pe^{-itH/n}P\right)^n f -
e^{-itK}f\|^2 dt = 0, \quad \mbox{for each} \; f \in \goth
\; \mbox{and} \; T > 0
\ee
holds. 
This result yields the existence of the limit of the Zeno product formula 
for almost all $t$ in the strong
operator topology, along a subsequence $\{n'\}$ of natural
numbers\footnote{This fact that the proof in \cite{EI} yields
convergence along a subsequence was omitted in the first version of the
paper from which the claim was reproduced in the review
\cite{Sch}. A complete proof of this claim is known at present only in
the case when $P$ is finite-dimensional, cf.~\cite{EI}.}. 

The reason why this result is weaker than the natural conjecture 
is that the exponential function involved in the
interlaced product gives rise to oscillations which are not easy
to deal with. One of the main ingredients in the present paper is
a simple observation that one can avoid the mentioned problem when
$\phi(x)= e^{-ix}$ is replaced by functions with an imaginary part
of constant sign. In analogy with the Kato class of the product formula
\eqref{0.2} it seems to be useful 
to introduce a class of admissible functions.
\bd\la{D1}
{\em
We call a Borel measurable function $\phi(\cdot): [0,\infty)
\longrightarrow \bC$ admissible if the conditions
\be\la{ass}
|\phi(x)| \le 1, \quad x \in [0,\infty),\; \quad 
\; \phi(0) = 1, \; \quad \mbox{and} \quad \phi'(+0) = -i,
\ee
are satisfied.
}
\ed
Typical examples are
\be
\phi(x) = (1+ix/k)^{-k}, \quad k = 1,2,\ldots, \quad \mbox{and} \quad 
\phi(x) = e^{-ix}, \quad x\in [0,\infty).
\ee
The main goal of this paper is to prove the following result.
\bt\la{II.6a}
Let $H$ be a non-negative self-adjoint operator in $\gotH$ and let
$\goth$ be a closed subspace of $\gotH$ such that $P: \gotH
\longrightarrow \goth$ is the
orthogonal projection from $\gotH$ onto $\goth$. If $\dom(\sqrt{H}) \cap
\goth$ is dense in $\goth$ and $\phi$ is admissible function  which obeys
\be\la{ass1}
\imag(\phi(x)) \le 0, \quad x \in [0,\infty),
\ee
then for any $t_0> 0$  one has
\be\la{gzeno}
\slim_{n\to\infty} \left(P\phi(tH/n)P\right)^n =
e^{-itK}\,, 
\ee
uniformly in $t \in [0,t_0]$, where the generator $K$ is defined
by (\ref{1.5}) and the strong convergence is meant on $\goth$.
\et
One may consider formul{\ae} of type \eqref{gzeno} as modified Zeno product
formul{\ae}. Examples of admissible functions obeying \eqref{ass1} are 
\bed
\phi(x) = (1+ix)^{-1} \quad \mbox{and} \quad \phi(x) = (1 +
ix/2)^{-2},  \quad x \in [0,\infty).
\eed
Unfortunately, not all admissible function do satisfy the condition
\eqref{ass1}. Indeed, the functions $\phi(x) = (1 + ix/3)^{-3}$ and $\phi(x) =
e^{-ix}$,  $x \in [0,\infty)$, are admissible but do not obey
\eqref{ass1}. In particular this yields that the convergence 
problem for the original Zeno product formula \eqref{0.3} is not
solved by Theorem \ref{II.6a} and remains open. 

However, Theorem \eqref{II.6a} suggests the following regularizing
procedure. We set
\bed
\gD_\phi := \{x \in [0,\infty): \imag(\phi(x)) \le 0\}.
\eed
By \eqref{ass} the set $\gD_\phi$ contains a neighbourhood of zero. If
the subset $\gD \subseteq \gD_\phi$ contains also a neighbourhood of zero, then
\bed
\phi_\gD(x) := \phi(x)\chi_\gD(x), \quad x \in [0,\infty),
\eed
defines an admissible function obeying $\imag(\phi_\gD(x)) \le 0$,
$x \in [0,\infty)$. By Theorem~\ref{II.6a} we obtain that for any $t_0 > 0$ one has
\bed
\slim_{n\to\infty}\left(P\phi_\gD(tH/n)P\right)^n = e^{-itK}
\eed
uniformly in $t \in [0,t_0]$. Applying this procedure to 
$\phi(x) = e^{-ix}$, $x \in [0,\infty)$, one has to choose a subset 
$\gD \subseteq \gD_\phi := \cup^\infty_{m=0} [2m\pi,(2m+1)\pi]$ containing
a neighbourhood of zero. 
>From  $\phi(x) = e^{-ix}$ one can construct a ``cutoff''
admissible function $\phi_\gD(x) :=
e^{-ix}\chi_{\gD}(x)$, $x \in [0,\infty)$, obeying \eqref{ass1}. 
In particular for  $\gD = [0,\pi)$, the function
\bed
\phi_\gD(x) = e^{-ix}\chi_\gD(x), \quad x \in [0,\infty),
\eed
is admissible and obeys \eqref{ass1}. This leads immediately to the following corollary.
\bc\la{II.6}
If the assumptions of Theorem \ref{II.6a} are satisfied, 
then for any $t_0> 0$ one has
\be\la{r1}
\slim_{n\to\infty} \left(P(I + itH/n)^{-1}P\right)^n =
\slim_{n\to\infty} \left(P(I + itH/2n)^{-2}P\right)^n =
e^{-itK}\,, 
\ee
and
\be\la{exp1}
\slim_{n\to\infty} \left(PE_H([0,\pi
n/t))e^{-itH/n}P\right)^n = e^{-itK} 
\ee
uniformly in $t \in [0,t_0]$ where $E_H(\cdot)$ is
the spectral measure of $H$, i.e. $H = \int_{[0,\infty)}
\gl\;dE_H(\gl)$. 
\ec
The ideas to replace the unitary group $e^{-itH}$ by a resolvent, cf. \eqref{r1}, or to 
employ a spectral cut-off together with $e^{-itH}$, cf. \eqref{exp1},
are not new:
they were used to derive a modification of the unitary Lie-Trotter
formula in \cite{Ich} and \cite{La1,La2}, respectively, both for
the form sum of two non-negative self-adjoint operators. See also
\cite{Ca1}.

Finally, let us note that 
formula \eqref{exp1} admits a physical interpretation in the context
of the Zeno effect. To this end we
note that the combination of the energy filtering and
non-decay measurement following immediately one after another,
see \eqref{exp1},  can
be regarded as \emph{a single generalized measurement.} In fact, a
product of two, in general non-commuting\footnote{We are primarily
interested, of course, in the nontrivial case when the $P$ does
not commute with $H$, and thus also with the spectral projections
$E_H([0,\pi n/t))$.} projections represents the simplest
non-trivial example of generalized observables\footnote{Since the
spectral projections involved commute with the evolution operator,
one can also replace the product $PE_H([0,\pi t/n))$ in our
formul{\ae} by $E_H([0,\pi t/n))PE_H([0,\pi t/n))$. Such
generalized observables represented by symmetrized projection
products have been recently studied as \emph{almost sharp quantum
effects} -- cf.~\cite{AG}.} introduced by Davies which are realized as positive maps of the
respective space of density matrices \cite[Sec.~2.1]{Da1}. Thus
formula \eqref{exp1}  corresponds to a modified Zeno situation
with such generalized measurements, which depend on $n$ and tend
to the standard non-decay yes-no experiment as $n\to\infty$.

Let us describe briefly the contents of the paper.
Section 2 is completely devoted to the proof of Theorem \ref{II.6a}.
In Section 3 we handle the general case of admissible functions under the
stronger assumption $\goth \subseteq \dom(\sqrt{H})$. We show that
under this assumption the modified Zeno product formula converges to 
$e^{-itK}$ for any admissible function not necessary satisfying the
additional condition \eqref{ass1}.  In particular, one has
\be\la{zeno}
\slim\left(Pe^{-itH/n}P\right)^n = e^{-itK}, 
\ee
uniformly in $t \in [0,t_0]$ for any $t_0 > 0$.
Moreover, we shall demonstrate there that 
under stronger assumptions, unfortunately too restrictive from the
viewpoint of physical applications, 
even the operator-norm convergence can be obtained. We finish the
paper with a conjecture which takes into account the results of \cite{EI} and the present
paper. 

\section{Proof of Theorem \ref{II.6a}}

We set
\be\la{1.1} 
F(\gt) := P\phi(\gt H)P: \goth \longrightarrow
\goth\,, \quad \gt \ge 0\,, 
\ee
and
\be\la{1.2} 
S(\gt) := \frac{I_\goth - F(\gt)}{\gt}: \goth
\longrightarrow \goth\,, \quad \gt > 0\,, 
\ee
where $I_\goth$ is the identity operator in the subspace $\goth$. 
In the following for an operator $X$ in $\gotH$
we use the notation $PXP$ for the operator $PXP :=
PX\!\upharpoonright\!\goth: \goth \longrightarrow \goth$ as well as for
its extension by zero in $\goth^\perp$.  
Let us assume that
\be\la{1.3}
\dom(T) := \dom(\sqrt{H}) \cap \goth
\ee
is dense in $\goth$. We define a linear operator $T: \goth
\longrightarrow \gotH$ by
\be\la{1.4}
Tf := \sqrt{H}f, \quad f \in \dom(T).
\ee
Since $\sqrt{H}$ is closed and $\dom(\sqrt{H}) \cap \goth$ is dense
the operator $T$ is closed and  its domain
$\dom(T)$ is dense in $\goth$. Then $T^*T: \goth \longrightarrow \goth$ is a
self-adjoint operator which is identical with $K$ defined by \eqref{1.5}, i.e. 
\be\la{1.4a}
K := T^*T: \goth \longrightarrow \goth
\ee
which defines a non-negative self-adjoint operator in $\goth$.

Further, let us represent the function $\phi$ as
\bed
\phi(x) = \psi(x) - i\go(x), \quad x \in [0,\infty),
\eed
where $\psi,\go: [0,\infty) \longrightarrow \bR$ are real-valued, Borel measurable functions obeying
\be\la{f}
|\psi(x)| \le 1, \quad \psi(0) = 1, \quad \psi'(+0) = 0
\ee
and
\bed
0 \le \go(x) \le 1, \quad \go(0) = 0, \quad \go'(+0) = 1.
\eed
Setting
\bed
\varphi(x) := 1 - \go(x), \quad x \in [0,\infty),
\eed
one has 
\be\la{g}
0 \le \varphi(x) \le 1,   \quad \varphi(0) = 1, \quad \varphi'(+0) = -1,
\ee
which shows that $\varphi$ is a Kato function. In terms of $\psi,\varphi$ the
function $\phi$ admits the representation
\bed
\phi(x) = \psi(x) - i(1 - \varphi(x)), \quad x \in [0,\infty).
\eed
We set
\be\la{p.1}
\ba{l}
p_-(x) := \left\{
\ba{lll}
1, & x = 0, & \\
\inf_{s \in (0,x]}(1 - \varphi(s))/s, & x > 0, & \mbox{and}
\ea
\right.\\[2mm]
p_+(x) := \left\{
\ba{ll}
1, & x = 0,\\
\sup_{s \in (0,x]}(1 - \varphi(s))/s, & x > 0.
\ea
\right.
\ea
\ee
Both functions are bounded on $[0,\infty)$ and obey
\be\la{p.2}
0 \le p_-(x) \le 1 \le p_+(x) < \infty,
\quad x \in [0,\infty).
\ee
The function $p_-$ is decreasing, i.e. 
$p_-(x) \ge p_-(y)$, $0 \le x \le y$, 
and $p_+$ is increasing, i.e. 
$p_-(x) \le p_-(y)$, $0 \le x \le y$.
We define the sesquilinear forms
\bed\la{p.3}
\gotk^-_\gt(f,g) := (p_-(\gt H)\sqrt{H}f,\sqrt{H}g), \quad f,g \in
\dom(\gotk^-_\gt) := \dom(\sqrt{H}) \cap \goth, \quad \gt \ge 0,
\eed
and
\bed\la{p.4}
\gotk^+_\gt(f,g) := (p_+(\gt H)\sqrt{H}f,\sqrt{H}g), \quad f,g \in
\dom(\gotk^+_\gt) := \dom(\sqrt{H}) \cap \goth, \quad \gt \ge 0.
\eed
Notice that for $\gt = 0$ one has $\gotk^-_0 = \gotk^+_0 = \gotk$ where the sesquilinear
form $\gotk$ is defined by  \eqref{1.5}. Obviously, both
forms $\gotk^\pm_\gt$ are non-negative for each $\gt \ge
0$. Moreover, the form $\gotk^-_\gt$ is closable for each $\gt > 0$
and its closure is a bounded form on $\goth$ while the form
$\gotk^+_\gt$ is already closed for each $\gt \ge 0$. 
By $K^\pm_\gt$ we denote the associated non-negative self-adjoint
operators on $\goth$. We note that $K^\pm_0 = K$. By \eqref{p.2} we get
\bed\la{p.5}
\gotk^-_\gt(f,f) \le \gotk(f,f) \le \gotk^+_\gt(f,f), \quad f \in
\dom(\gotk^-_\gt) = \dom(\gotk) = \dom(\gotk^+_\gt), \quad \gt \ge 0,
\eed
which yields
\bed\la{p.6}
K^-_\gt \le K \le K^+_\gt, \quad \gt \ge 0.
\eed
Since $p_-$ is decreasing the family $\{K^-_\gt\}_{\gt \ge 0}$ is
increasing as $\gt \downarrow 0$. Further, from \eqref{p.1} one gets that $\slim_{\gt\to+0}p_-(\gt H) = I_\gotH$.
Since $\gotk^-_\gt \le \gotk$ and
\bed
\lim_{\gt\to+0}\gotk^-_\gt(f,g) = \lim_{\gt\to+0}(p_-(\gt H)\sqrt{H}f,\sqrt{H}g)
= \gotk(f,g), 
\eed
$f,g \in \dom(\gotk^-_\gt) = \dom(\gotk)$, we obtain from Theorem VIII.3.13 of \cite{Ka} that
\be\la{p.7}
\slim_{\gt\to+0}(I_\goth + K^-_\gt)^{-1} = (I_\goth + K)^{-1}.
\ee
Further, since $p_+$ is increasing the family $\{K^+_\gt\}_{\gt \ge
  0}$ is decreasing as $\gt \downarrow 0$.
By $\slim_{\gt\to+0}p_+(\gt H) = I_\gotH$ we find
\bed
\lim_{\gt\to+0}\gotk^+_\gt(f,g) = \lim_{\gt\to+0}(p_+(\gt H)\sqrt{H}f,\sqrt{H}g)
= \gotk(f,g),
\eed
$f,g \in \dom(\gotk^+_\gt) = \dom(\gotk)$.
Since $\gotk$ is closed we obtain from Theorem VIII.3.11 of \cite{Ka}
that
\be\la{p.8}
\slim_{\gt\to+0}(I_\goth + K^+_\gt)^{-1} = (I_\goth + K)^{-1}.
\ee
\bl\la{II.1a}
Let $\left\{X(\gt)\right\}_{\gt > 0}$, $\left\{Y(\gt)\right\}_{\gt
> 0}$, and $\left\{A(\gt)\right\}_{\gt > 0}$ be families of bounded
non-negative self-adjoint operators in $\goth$ such that the
condition
\bed 0 \le X(\gt) \le A(\gt) \le Y(\gt)\,, \quad \gt > 0\,, \eed
is satisfied. If  $s\!-\!\lim_{\gt \to 0}X(\gt) = s\!-\!\lim_{\gt \to 0}Y(\gt) =
A$, where $A$ is a bounded self-adjoint operator in $\goth$, then
$s\!-\!\lim_{\gt \to 0}A(\gt) = A$.
\el
\begin{proof}
Since for each $f \in \goth$ we have
\bed (X(\gt)f,f) \le (A(\gt)f,f) \le (Y(\gt)f,f)\,, \quad \gt >
0\,, \eed
we get $\lim_{\gt\to 0}(A(\gt)f,f) = (Af,f)$, $f \in \goth$, or
$w\!-\!\lim_{\gt\to 0}A(\gt) = A$. Hence 
\bed
w-\lim_{\gt\to 0} (Y(\gt) - A(\gt)) = 0. 
\eed
Since $Y(\gt) - A(\gt) \ge 0$, $\gt > 0$, we find
\bed
\slim_{\gt\to 0} (Y(\gt) - A(\gt))^{1/2} = 0 
\eed
which yields $\slim_{\gt\to 0} (Y(\gt) - A(\gt)) = 0$. Hence
$\slim_{\gt\to 0} A(\gt) = A$.
\end{proof}
>From \eqref{1.2} we obtain
\be\la{a0}
S(\gt) = \frac{1}{\gt}P(I_\gotH - \psi(\gt H))P + 
i\frac{1}{\gt}P(I_\gotH - \varphi(\gt H))P, \quad \gt > 0.
\ee
Let
\be\la{b0}
L_0(\gt) := \frac{1}{\gt}P(I_\gotH - \varphi(\gt H))P: \goth \longrightarrow
\goth, \quad \gt > 0.
\ee
\bl\la{III.1}
Let $H$ be a non-negative self-adjoint operator in $\gotH$ and let
$\goth$ be a closed subspace of $\gotH$. If $\dom(\sqrt{H}) \cap \goth$ is
dense in $\goth$ and $\varphi$ is a Kato function, then we have
\be\la{3.00} 
\slim_{\gt\to 0}\left(I_\goth + L_0(\gt)\right)^{-1} = (I_\goth + K)^{-1}
\ee
\el
\begin{proof}
Since
\bed
\gotk^-_\gt(f,f) \le \left(\frac{I_\gotH - \varphi(\gt H)}{\gt}f,f\right) \le \gotk^+_\gt(f,f),
\quad f \in \dom(\sqrt{H}) \cap \goth,
\eed
we find
\bed
K^-_\gt \le P\frac{I_\gotH - \varphi(\gt H)}{\gt}P \le K^+_\gt, \quad
\gt > 0.
\eed
Hence
\bed
X(\gt) := (I_\goth + K^+_\gt)^{-1} \le 
\left(I_\goth +  P\frac{I_\gotH - \varphi(\gt H)}{\gt}P\right)^{-1}
\le (I_\goth + K^-_\gt)^{-1} =: Y(\gt),
\eed
$\gt > 0$. Taking into account \eqref{p.7},\eqref{p.8} and applying Lemma
\ref{II.1a} we prove \eqref{3.00}.
\end{proof}
We set
\bed
L(\gt) := \frac{1}{\gt}P(I  - \psi(\gt H))P + 
\frac{1}{\gt}P(I - \varphi(\gt H))P : \goth \longrightarrow \goth, 
\quad \gt > 0.
\eed
\bl\la{II.5}
Let $H$ be a non-negative self-adjoint operator in $\gotH$ and
let $\goth$ be  a closed subspace of $\gotH$. If $\dom(\sqrt{H}) \cap
\goth$ is dense in $\goth$, the real-valued Borel measurable
function $\psi$ obeys \eqref{f} and $\varphi$ is a Kato function, then
\be\la{1.19} 
\slim_{\gt\to 0} (I_\goth + L(\gt))^{-1} = (I_\goth + K)^{-1}. 
\ee
\el
\begin{proof}
Let
\bed
\zeta(x) := \frac{\psi(2x) + \varphi(2x)}{2}, \quad x \in [0,\infty).
\eed
Notice that $\zeta$ is a Kato function. Setting
\bed
\widetilde{L}_0(\gt) := \frac{1}{\gt}P(I_\gotH - \zeta(\gt H))P, \quad \gt
> 0,
\eed
we obtain from Lemma \ref{III.1} that
$\slim_{\gt\to 0} = (I_\goth + \widetilde{L}_0(\gt))^{-1} = (I_\goth + K)^{-1}$.
By $L(2\gt) = \widetilde{L}_0(\gt)$ we prove \eqref{1.19}.
\end{proof}
We set
\bed
M(\gt) := (I_\goth + L_0(\gt))^{-1/2}P\frac{I - \psi(\gt H)}{\gt}P(I_\goth + L_0(\gt))^{-1/2},
\quad \gt > 0.
\eed
\bl
Let $H$ be a non-negative self-adjoint operator in $\gotH$ and let
$\goth$ be a closed subspace of $\gotH$. If $\dom(\sqrt{H}) \cap \goth$ is
dense in $\goth$, the real-valued, Borel measurable functions $\psi$
obeys \eqref{f} and $\varphi$ is a Kato function, then we have
\be\la{b1}
\slim_{\gt\to 0}\left(I_\goth + M(\gt)\right)^{-1} = I_\goth.
\ee
\el
\begin{proof}
A straightforward computation proves the representation
\bed
(I_\goth + L(\gt))^{-1} = (I_\goth + L_0(\gt))^{-1/2}\left(I_\goth + M(\gt)\right)^{-1}(I_\goth + L_0(\gt))^{-1/2},
\quad \gt > 0.
\eed 
By \eqref{3.00} and \eqref{1.19} we get
\bed
w-\lim_{\gt\to 0}\left(I_\goth + M(\gt)\right)^{-1} = I_\goth
\eed
which yields
\bed
\slim_{\gt\to 0}\left(I_\goth -\left(I_\goth + M(\gt)\right)^{-1}\right)^{1/2} = 0.
\eed
Hence
\bed
\slim_{\gt\to 0}\left(I_\goth - \left(I_\goth + M(\gt)\right)^{-1}\right) = 0
\eed
which proves \eqref{b1}.
\end{proof}
>From \eqref{b1} one gets
\bed
\slim_{\gt\to 0}\left(iI_\goth + M(\gt)\right)^{-1} = -iI_\goth.
\eed
Hence
\bed
\slim_{\gt\to 0}(I_\goth + L_0(\gt))^{-1/2}\left(iI_\goth + M(\gt)\right)^{-1}(I_\goth + L_0(\gt))^{-1/2} 
= -i(I_\goth + K)^{-1}.
\eed
or
\bed
\slim_{\gt\to 0}\left(iI_\goth + \frac{1}{\gt}P(I_\goth - \psi(\gt H))P + iL_0(\gt)\right)^{-1}
= (iI_\goth + iK)^{-1}.
\eed
Using \eqref{a0} and\eqref{b0} we obtain
\bed
\slim_{\gt\to 0}(iI_\goth + S(\gt))^{-1} = (iI_\goth + iK)^{-1}
\eed
which yields
\bed
\slim_{\gt\to 0}(I_\goth + S(\gt))^{-1} = (I_\goth + iK)^{-1}
\eed
We finish the proof of Theorem \ref{II.6a} applying  Chernoff's theorem \cite{Ch}
or Lemma~3.29 of \cite{Da2}.

\section{Arbitrary admissible functions} \label{s: norm}

Theorem \ref{II.6a} needs the additional assumption \eqref{ass1} and
it is unclear whether this assumption can be dropped. In the following
we are going to show that under stronger assumptions on the domain of
$\sqrt{H}$ the condition \eqref{ass1} is indeed not necessary.
\bt\la{IV.1}
Let $H$ be a non-negative self-adjoint operator on
$\gotH$ and let $\goth$ be a closed subspace of
$\gotH$ such that $P: \gotH \longrightarrow \goth$ is the orthogonal
projection from $\gotH$ onto $\goth$. If $\goth \subseteq
\dom(\sqrt{H})$ and $\phi$ is admissible, then
\be\la{4.1} \slim_{n\to\infty} \left(P\phi(tH/n)P\right)^n =
e^{-itK} \ee
uniformly in $t \in [0,t_0]$ for any $t_0 > 0$ where $K$ is
defined by (\ref{1.5}).
\et
\begin{proof}
We note that $\goth \subseteq \dom(\sqrt{H})$ implies
that $T = \sqrt{H}P$ is a bounded operator, and consequently, $K =
T^*T$ is also bounded. We may employ the representation
\be\la{4.1e}
\left(\frac{I_\gotH - \phi(\gt H)}{\gt}f,g\right) = \left(p(\gt H)\sqrt{H}f,\sqrt{H}g\right),
\quad \gt > 0,
\ee
for $f \in \dom(H)$ and $g \in \dom(\sqrt{H})$ where
\bed
p(x) := \left\{
\ba{ll}
i, & x = 0\\
(1-\phi(x))/x, & x > 0
\ea
\right.\;.
\eed
Since $C_p := \sup_{x \in [0,\infty)}|p(x)| < \infty$ by \eqref{ass}
one gets $\|p(\gt H)\|_{\kB(\gotH)} \le C_p$, $\gt > 0$. Hence 
the equality \eqref{4.1e} extends to $f,g \in
\dom(\sqrt{H})$, in particular, to $f,g \in \goth$. This leads to the representation
\be\la{4.1a} 
(I_\goth - F(\gt))f = T^*p(\gt H)Tf\,, \quad \gt > 0\,, \quad f \in \goth\,, 
\ee
or
\be\la{4.1b}
S(\gt)f - iKf = 
T^*\left(p(\gt H) - iI_\gotH\right)Tf\,, \quad \gt > 0\,, \quad f
\in \goth\,.
\ee
By assumption \eqref{ass} we find
$\slim_{\gt\to 0}p(\gt H)  = iI_\gotH$ which yields 
$\slim_{\gt\to 0} S(\gt) = iK$.
In this way we obtain the relation
\bed 
\slim_{\gt\to 0}(I_\goth + S(\gt))^{-1} = (I_\goth + iK)^{-1}, 
\eed
and using Chernoff's theorem \cite{Ch} one more time we have proved
(\ref{4.1}).
\end{proof}
It turns out that the convergence \eqref{4.1} can be improved to
operator-norm convergence under some stronger assumption.
%
\bc\la{IV.2}
Let the assumptions of Theorem \ref{IV.1} be satisfied.
One has
\be\la{4.2}
\lim_{n\to\infty} \left\|\left(P\phi(tH/n)P\right)^n
-e^{-itK}\right\|_{\kB(\goth)} = 0
\ee
uniformly in $t \in [0,t_0]$ for any $t_0 > 0$ if in addition
\begin{enumerate}

\item[$\mbox{(i)}$]  the operator $T$ is compact or

\item[$\mbox{(ii)}$] there is $\ga > 0$ such that
$\goth \subseteq \dom(\sqrt{H^{1+\ga}})$ and $C_\ga := \sup_{x \in
  (0,\infty)}|p_\ga(x)| < \infty$ where
\bed
p_\ga(x) := 
\left\{
\ba{ll}
0, & x = 0\\
(p(x) - i)/x^{\ga}, & x > 0
\ea
\right.\;.
\eed
\end{enumerate}
\ec
\begin{proof}
>From \eqref{4.1b} and the compactness of $T$ we find 
\be\la{4.4}
\lim_{\gt \to 0}\left\|S(\gt) - iK\right\|_{\kB(\goth)} = 0.
\ee
If $\goth \subseteq \dom(\sqrt{H^{1+\ga}})$ for some $\ga > 0$,
then we set $T_\ga := \sqrt{H^{1+\ga}} P$ and $K_\ga = T^*_\ga T_\ga$. Notice
that $T_\ga$ is a bounded operator. From \eqref{4.1b} we obtain the
representation
\bed 
S(\gt) - iK = \gt^\ga \; T^*_\ga \;p_\ga(\gt H) \;T_\ga\;, \quad \gt > 0.
\eed
Hence we find the estimate
\bed 
\left\|S(\gt) - iK\right\|_{\kB(\goth)} \le
\gt^\ga C_\ga\|K_\ga\|_{\kB(\goth)},
\quad \gt > 0,
\eed
which yields \eqref{4.4}. Using the representation
\bed
e^{-itK} - e^{-tS(t/n)} = \int^t_0 e^{-(t-s)S(t/n)}(S(t/n) -
iK)e^{-isK} ds
\eed
we get the estimate
\bed
\left\|e^{-itK} - e^{-tS(t/n)}\right\|_{\kB(\goth)} \le t\|S(t/n) -
  iK\|_{\kB(\goth)}, \quad t \ge  0.
\eed
Using  \eqref{4.4} we find
\be\la{4.5}
\lim_{n\to\infty}\left\|e^{-tS(t/n)} - e^{-itK}\right\|_{\kB(\goth)}
=0
\ee
holds for any $t > 0$, uniformly in $t \in [0,t_0]$. We shall
combine it with the telescopic estimate
\bea\la{4.5a}
\lefteqn{
\left\|F(t/n)^n - e^{-itK}\right\|_{\kB(\goth)} \le}\\
& & \le \left\|F(t/n)^n - e^{-tS(t/n)}\right\|_{\kB(\goth)} +
\left\|e^{-tS(t/n)} - e^{-itK}\right\|_{\kB(\goth)}\,,\nonumber
\eea
where the first term can be treated as in Lemma 2 of \cite{Ch1}, see
also \cite[ Lemma 3.27]{Da2},
\be\la{4.6} \left\|\left(F(t/n)^n -
e^{-tS(t/n)}\right)f\right\| \le
\sqrt{n}\left\|\left(F(t/n) - I_\goth\right)f\right\|\,,
\quad f \in \goth\,. \ee
Using the representation (\ref{4.1a}) with $\gt = t/n$, we can
estimate the right-hand side of \eqref{4.6} by
\bed 
\left\|\left(F(t/n) - I_\goth\right)f\right\| \le
\frac{t}{n}\left\|T^*p(tH/n)Tf\right\|\,, 
\quad f \in
\goth\,,\quad t > 0\;.
\eed
Since $\|p(\gt H)\|_{\kB(\gotH)} \le C_p$, $\gt > 0$,  we find
\bed 
\left\|\left(F(t/n) - I_\goth\right)f\right\| \le
C_p\frac{t}{n} \|K\|_{\kB(\goth)}\|f\|\,, \quad f \in \goth\,. \eed
Inserting this estimate into (\ref{4.6}) we obtain
\bed
\left\|F(t/n)^n - e^{-tS(t/n)}\right\|_{\kB(\goth)} \le
C_p\frac{t}{\sqrt{n}}\|K\|_{\kB(\goth)}
\eed
which yields
\be\la{4.8}
\lim_{n\to\infty}\left\|F(t/n)^n -
  e^{-tS(t/n)}\right\|_{\kB(\goth)} = 0
\ee
for any $t > 0$, uniformly in $t \in [0,t_0]$. Taking into account
(\ref{4.5}), (\ref{4.5a}) and (\ref{4.8}) we arrive at the sought
relation (\ref{4.2}).
\end{proof}
\begin{rem}
{\em
Since $\phi(x) = e^{-ix}$, $x \in [0,\infty)$, is admissible we get
from Theorem \ref{IV.1} that under the assumptions $\goth \subseteq
\dom(\sqrt{H})$ the original Zeno product formula \eqref{zeno} holds
and that under the stronger assumptions $\sqrt{H}P$ is compact or
$\goth \subseteq \dom(\sqrt{H^{1+\ga}})$, $\ga > 0$, the original Zeno product formula
\eqref{zeno} converges in the operator norm.
}
\end{rem}
\begin{rem}
{\em
Obviously, the conclusion (\ref{4.2}) is valid if $\goth \subseteq
\dom(\sqrt{H})$ and $\goth$ is a finite dimensional subspace.
Indeed, in this case the operator $T$ is finite dimensional, and
therefore compact. This gives an alternative proof of the result
derived in Section~5 of \cite{EI} for the case $\phi(x) = e^{-ix}$.
}
\end{rem}
\begin{rem}
{\em
In connection with the previous remark let us  mention that in
the finite-dimensional case there is one more way to prove the
claim suggested by G.M.~Graf and A.~Guekos \cite{GG} for the special
case $\phi(x) = e^{-ix}$. The argument
is based on the observation that
\be\la{4.99} \lim_{t\to 0}\: t^{-1}\left\| P e^{-itH}P - P
e^{-itK}P \right\|_{\kB(\goth)} = 0 \ee
implies $\left\| (P e^{-itH/n}P)^n - e^{-itK}
\right\|_{\kB(\goth)} = n\, o(t/n)$ as $n\to\infty$ by means of a
natural telescopic estimate. To establish (\ref{4.99}) one first
proves that
\bed t^{-1} \left[ (f, P e^{-itH}Pg) -
(f,g) -it (\sqrt{H}Pf, \sqrt{H}Pg)
\right] \longrightarrow 0 \eed
as $t\to 0$ for all $f,g$ from $\dom(\sqrt{H}P)$ which coincides
in this case with $\goth$ by assumption. The last
expression is equal to
\bed \left( \sqrt{H}Pf, \left[ \frac{e^{-itH}-I}{tH} -i \right]
\sqrt{H}Pg \right) \eed
and the square bracket tends to zero strongly by the functional
calculus, which yields the sought conclusion. We note that the
operator in the square brackets is well-defined by the functional
calculus even if $H$ is not invertible. In the same way we
find that
\bed t^{-1} \left[ (f, P e^{-itK}Pg) -
(f,g) -it (\sqrt{K}f, \sqrt{K}g) \right]
\longrightarrow 0 \eed
holds as $t\to 0$ for any vectors $f,g\in\goth$. Next we note that
$(\sqrt{K}f, \sqrt{K}g) = (\sqrt{H}Pf,
\sqrt{H}Pg)$, and consequently, the expression
contained in (\ref{4.99}) tends to zero weakly as $t\to 0$,
however, in a finite dimensional $\goth$ the weak and
operator-norm topologies are equivalent.
}
\end{rem}
\begin{conj}
{\em
Comparing the results of the present paper with those ones of
\cite{EI} we conjecture that if we drop the assumption
\eqref{ass1}  in Theorem \ref{II.6a}, then at least the convergence
\be\la{conj}
\lim_{n\to\infty}\int^T_0\left\|\left(\phi(tH/n)\right)^nf -
  e^{-itK}f\right\|^2\;dt = 0
\ee
holds for for each $f \in \goth$, $T > 0$ and arbitrary admissible
functions $\phi$. The proofs of \cite{EI} rely heavily on the 
analytic properties of the exponential function
$\phi(x) = e^{-ix}$. For admissible functions analytic
properties are not required which yields the necessity  to look for a
different proof idea.
}
\end{conj}

\section*{Acknowledgement}
This work was supported by Czech Academy of Sciences within the
project K1010104 and the ASCR-CNRS exchange program, and by
Ministry of Education of the Czech Republic within the project
LC06002 and the French-Czech CNRS bilateral project.
V. Z. and H. N. thanks the  Department of Theoretical Physics of the
Czech Academy of Sciences in \v Re\v z for hospitably
and financial support. P. E.  and H. N. are grateful to the Universit\'{e} de la M\'{e}diterran\'{e}e and
Centre de Physique Th\'{e}orique-CNRS-Luminy, Marseille (France), for 
hospitality extended to them and financial support. V. Z. is also
thankful the Weierstrass-Institut of Berlin for hospitality and
financial support in 2005 when the paper reached its final form. 

\end{document}